\documentclass[aps,prb,twocolumn,superscriptaddress,showpacs]{revtex4}
\usepackage{bm}
\usepackage{epsf}
\usepackage{amssymb}
\usepackage{amsmath}
\usepackage{graphicx}
\usepackage{rotating}
\usepackage{epsfig}
\usepackage{psfrag}
\usepackage{amsmath}
\usepackage[breaklinks]{hyperref}
\usepackage{hyperref}

\hypersetup{
    bookmarks=true,         
    unicode=false,          
    pdftoolbar=true,        
    pdfmenubar=true,        
    pdffitwindow=true,      
    pdftitle={My title},    
    pdfauthor={Author},     
    pdfsubject={Subject},   
    pdfcreator={Creator},   
    pdfproducer={Producer}, 
    pdfkeywords={keywords}, 
    pdfnewwindow=true,      
    colorlinks=true,       
    linkcolor=red,          
    citecolor=blue,        
    filecolor=magenta,      
    urlcolor=blue           
}

\DeclareMathAlphabet{\bi}{OML}{cmm}{b}{it}

\begin{document}
\def\G{{\cal G}}
\def\F{{\cal F}}
\def\ea{\textit{et al.}}
\def\bM{{\bm M}}
\def\bN{{\bm N}}
\def\bV{{\bm V}}
\def\bj{\bm{j}}
\def\bSig{{\bm \Sigma}}
\def\bLam{{\bm \Lambda}}
\def\bfeta{{\bf \eta}}
\def\bn{{\bf n}}
\def\d{{\bf d}}
\def \xy{$x$--$y$ }
\def\bP{{\bf P}}
\def\bK{{\bf K}}
\def\bk{{\bf k}}
\def\bkn{{\bf k}_{0}}
\def\bx{{\bf x}}
\def\bz{{\bf z}}
\def\bR{{\bf R}}
\def\br{{\bf r}}
\def\bu{{\bm u}}
\def\bq{{\bf q}}
\def\bp{{\bf p}}
\def\by{{\bf y}}
\def\bQ{{\bf Q}}
\def\bs{{\bf s}}
\def\bA{{\mathbf A}}
\def\bv{{\bf v}}
\def\b0{{\bf 0}}
\def\la{\langle}
\def\ra{\rangle}
\def\Im{\mathrm {Im}\;}
\def\Re{\mathrm {Re}\;}
\def\beq{\begin{equation}}
\def\eeq{\end{equation}}
\def\bdm{\begin{displaymath}}
\def\edm{\end{displaymath}}
\def\bnab{{\bm \nabla}}
\def\Tr{{\mathrm{Tr}}}
\def\bJ{{\bf J}}
\def\bU{{\bf U}}
\def\bPsi{{\bm \deltaDelta}}
\def\mA {\mathrm{A}}
\def \R{R_{\mathrm{s}}}
\def \rhos{n_{\mathrm{s}}}
\def \rhon{\tilde{n}}
\def \Rd{R_{\mathrm{d}}}
\def \xy{three dimensional $XY\;$}
\def\sfrac{\textstyle\frac}
\def\e0{\epsilon_B}
\def\ath{a_{3}}
\def\atw{a_{2}}
\def\ntw{n_{\mathrm{2D}}}
\def\2d{2d}
\def\3d{3d}
\def\gtw{\ln(k_F\atw)}
\def\gth{-1/(k_F\ath)}
\def\A{\bar{A}}
\def\bG{\hat{{\bm G}}}
\def\bGt{\hat{\tilde{{\bm G}}}}
\def\Sr{\mathrm{Sr}_2\mathrm{RuO}_4}
\def\cG{c_{\mathrm{G}}}
\def\cM{c_{\mathrm{M}}}
\def\HeB{$^3$He-$B$ }
\def\HeBp{$^3$He-$B$}
\def\bH{{\bf H}}
\def\Px{{\hat{\sigma}_x}}
\def\Py{{\hat{\sigma}_y}}
\def\Pz{{\hat{\sigma}_z}}
\def\Pj{{\hat{\sigma}_j}}

\title{Locally gauge-invariant spin response of $^3$He-$B$ films with Majorana surface states}
\author{Edward~Taylor}
\affiliation{Department of Physics and Astronomy, McMaster University, Hamilton, Ontario, L8S 4M1, Canada}
\author{A. John Berlinsky}
\affiliation{Kavli Institute for Theoretical Physics, University of California, Santa Barbara, California, 93106-9530, USA}
\author{Catherine~Kallin}
\affiliation{Department of Physics and Astronomy, McMaster University, Hamilton, Ontario, L8S 4M1, Canada}
\affiliation{Kavli Institute for Theoretical Physics, University of California, Santa Barbara, California, 93106-9530, USA}
\affiliation{Canadian Institute for Advanced Research, Toronto, Ontario M5G 1Z8, Canada}
\date{\today}
\begin{abstract}
A locally gauge-invariant theory of the spin response of a thin film of $^3$He-$B$ film is given that describes fluctuation effects arising from the coupled dynamics of the superconducting order parameter (the collective mode) and in-gap Majorana surface states.  In contrast to a mean-field calculation of the spin response, which predicts a nonzero imaginary longitudinal spin susceptibility at frequencies inside the bulk gap due to absorption from the Majorana states, our gauge-invariant theory shows that this response is strongly suppressed above the collective mode frequency and vanishes if dipole-dipole interactions are neglected.   In the presence of dipole-dipole interactions, in sufficiently thin films, and at ultra-low temperatures, the Majorana states lead to a distinctive magnetic-field- and temperature-dependent damping of the collective mode, a feature that may be observable in longitudinal NMR experiments. 
\end{abstract}

\maketitle

\section{Introduction}
An extremely active sub-branch of condensed matter physics has concerned itself over the past decade with the quest to find Majorana bound states in topological insulators~\cite{Hasan10} and superconductors~\cite{Beenakker13}.  These are localized zero-energy excitations that obey non-Abelian statistics, meaning that a pair of them widely separated in space can be prepared in a superposition which is protected by topology against decoherence.  

There are in principle several ways one can go about looking for Majorana bound states in topological superconductors.  The ``gold standard'' is unquestionably an interference-type measurement, which directly probes their non-Abelian nature.  These are difficult, however, and an appealing second option for experimentalists is to look for signatures of the zero-energy nature of Majorana states.  Here there are essentially two possibilities: probe the single-particle Green's function via surface density-of-states (DOS) measurements~\cite{Law09,Flensberg10,Mourik12} or look for signatures in two-particle response functions.  While the former is a seemingly direct probe of the zero-energy nature of Majorana bound states,  zero-bias anomalies in the DOS arise in a number of contexts, and do not necessarily imply the existence of Majoranas.  (At the same time, indirect probes of the DOS of superfluid  $^3$He, including transverse acoustic impedance~\cite{Murakawa09,Murakawa11}, specific heat~\cite{Choi06}, and sound attenuation~\cite{Davis08} measurements, are consistent with a gapless spectrum within the bulk superfluid gap.)  The latter approach, on the other hand, seems ideally suited for charge-neutral topological superfluids such as $^3$He, for which surface DOS measurements are not possible.

There is of course a precedent for looking for signatures of zero-energy modes in two-particle response functions.  The DC Hall conductivity in two-dimensional electron gases (2DEGs) is exactly equal to fundamental constants multiplied by the Chern number~\cite{Niu85}, a topological invariant that provides a measure of the number of zero-energy modes along the edge of this system.    Unlike 2DEGs, however, the low-frequency response properties of superconductors are dominated by the appearance of a gapless Goldstone branch~\cite{Anderson58}.  (The fact that the physical Goldstone mode is gapped out in electronic systems does not matter: the zero-energy poles of the response functions survive polarization corrections~\cite{Martin,Schriefferbook}.)   Given the presence of \emph{two} in-gap excitation branches---Majorana and Goldstone---in a topological superconductor, it is natural to ask whether signatures of the former are obfuscated by the latter, especially given that the poles of two particle response functions coincide with the collective mode (Goldstone) spectrum~\cite{NozieresPines}.  

As an example of this, the interplay of the gapless Majorana edge and Goldstone branches  in a two-dimensional time-reversal symmetry-breaking chiral $p$-wave superconductor leads to a non-local optical Hall conductivity~\cite{Goryo98,Roy08}
\beq \sigma_{xy}(\omega,\bq) = \frac{\tilde{C}}{4\pi} \frac{c^2q^2}{c^2q^2-\omega^2}.\label{sigmaHall}\eeq
Here $\tilde{C}$ reduces to the Chern number in the limit where the superconducting gap $\Delta_0$ is much smaller than the Fermi energy $E_F$ and $c \simeq v_F/\sqrt{2}$ is the sound velocity in this limit.  Equation~(\ref{sigmaHall}) shows that the optical Hall response of a chiral superconductor is dominated by the Goldstone pole at $\omega=cq$.  The only hope for detecting a Majorana branch (via a nonzero Chern number) is to probe \emph{below} the Goldstone pole, with a spatially inhomogeneous electric field such that $\omega\ll cq$, a practical impossibility given that the the speed of light is generally much larger than the speed of sound.  
 
 In this paper, we turn our attention to another manifestation of this same physics and consider the longitudinal spin susceptibility of a thin film of $^3$He-$B$ which is predicted to harbour Majorana states at its surfaces.  Recently, a number of authors have argued that these states would give rise to a spectral  feature in the spin susceptibility which could be observed in nuclear magnetic resonance (NMR) experiments~\cite{Chung09,Nagato09,Silaev11,Tsutsumi11,Mizushima12}.  These predictions are based on mean-field calculations, however, and ignore effects related to the collective modes which arise from order parameter fluctuations.  Here we derive a locally gauge-invariant theory of the spin response function.  As with the optical Hall response (\ref{sigmaHall}), we find that the spin response is dominated by the collective modes of the superconducting order parameter---in this case, longitudinal spin modes propagating in the plane of the film.  Signatures of the Majorana states manifest themselves in the damping of these spin modes.  Extending an idea put forth by Silaev~\cite{Silaev11}, a fingerprint of Majorana states can be found in the tunability of this damping in the presence of a magnetic field.  

We start in Sec.~\ref{meanfieldsec} by reviewing the mean-field theory of Majorana surface states in $^3$He-$B$ and the resulting spin susceptibility.  In Sec.~\ref{spinmodesec}, we derive an action describing the low-energy longitudinal spin modes for this system.  We use this action in Sec.~\ref{GIspinresponse} to derive the locally gauge-invariant longitudinal spin susceptibility.  Finally, in a concluding Sec.~\ref{conclusions}, we discuss the prospects for observing signatures of Majorana bound states in NMR.  

\section{Surface states in a thin film of $^3$He-$B$ and the mean-field spin susceptibility}
\label{meanfieldsec}
\begin{center}
\begin{figure}
\includegraphics[width=0.8\linewidth]{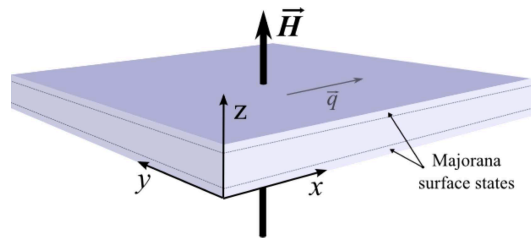}
\caption{Experimental configuration to measure the signature of Majorana bound states using NMR.  A thin film of $^3$He-$B$ is confined to a slab above $z=0$ and subjected to a magnetic field $\bH = \hat{\mathbf{z}}H_z$.  Low-energy Majorana surface states disperse in-plane [$\bk = (k_x,k_y)]$ and are localized to within a coherence length of the upper and lower surfaces. Collective modes--in this case, the longitudinal spin mode--arising from order parameter fluctuations propagate with wavevector $\bq$ in-plane as well.}\label{fig1}
\end{figure}
\end{center}

The spectrum of the surface states  of a thin film of superfluid $^4$He-$B$ has been calculated by a number of authors; see, e.g., Refs.~\onlinecite{Volovik09,Volovik10,Silaev11}.  Below we review the semiclassical solution of the relevant Bogoliubov-de Gennes (BdG) equations and use it to calculate the frequency-dependent mean-field spin susceptibility.  The experimental configuration we have in mind is shown in Fig.~\ref{fig1}: A thin film of superfluid $^3$He-$B$ confined to $0<z<h$, where $h$ is the film thickness, is subjected to an external magnetic field $\bH(\br,t) = \mathbf{\hat{z}}H_z(\br,t)$.  Majorana surface states are confined to within a coherence length $\xi_0$ of the lower surface at $z=0$ as well as the upper surface at $z=h$.  

To simplify calculations, we will carry out semiclassical calculations for a half-infinite space, bounded from below at $z=0$, removing the top surface from Fig.~\ref{fig1}.  If the film is very narrow, the Majorana bound states at one surface are sensitive to the existence of those at the other, leading to a splitting of the spectrum that is exponential in the film width~\cite{Tsutsumi11}:
\beq \Delta E \sim \Delta_0\exp(-h/3\xi_0).\label{splitting}\eeq  In this paper, we will only be interested in situations where this energy splitting is the smallest energy scale in the problem, meaning that we can consider a single surface and treat the two surfaces as independent, doubling our final result for the contribution of the Majorana states to the spin susceptibility.  Mindful of the actual thin-film geometry, we will however use a form for the order parameter appropriate for this geometry (see below). 

In the half-infinite geometry, using the spinor basis $\hat{\Psi}^{\dagger}\equiv (\psi^{\dagger}_{\uparrow},\psi^{\dagger}_{\downarrow},\psi_{\uparrow},\psi_{\downarrow})$ and for a magnetic field along $z$, the semiclassical BdG Hamiltonian is~\cite{Volovik09} 
\beq {\cal{H}} = \left(\begin{array}{cc} -iv_F\partial_z-\frac{\gamma}{2} H_z\hat{\sigma}_z & \hat{\Delta}(\bk)\\ \hat{\Delta}^{\dagger}(\bk) & iv_F\partial_z +\frac{\gamma}{2} H_z\hat{\sigma}_z\end{array}\right).\label{H}\eeq
Here $\hat{\sigma}_z$ is the Pauli (nuclear) spin matrix and $\gamma$ is the gyromagnetic ratio of $^3$He.
For $^3$He-$B$ confined to a thin film, dipolar forces constrain the form of the order parameter (see the appendix).  Within the same semiclassical approximation used to obtain (\ref{H}), it is 
\beq \hat{\Delta}(\bk)= \frac{\Delta_0}{k_F}\left(\begin{array}{cc} -k_x+ik_y & k_F\\ k_F & k_x+ik_y\end{array}\right).\label{hz}\eeq

Following Silaev~\cite{Silaev11} and Volovik~\cite{Volovik09}, we solve (\ref{H}) perturbatively by treating the magnetic field as small as well as the in-plane momentum $\bk\equiv (k_x, k_y)$ as compared to $k_F$.  We thus decompose (\ref{H}) as ${\cal{H}}= {\cal{H}}^{(0)} + {\cal{H}}^{(1)}$, where
\beq {\cal{H}}^{(0)} = \left(\begin{array}{cc}  -iv_F\partial_z & \Delta_{0}\Px \\ \Delta_{0}\Px & iv_F\partial_z \end{array}\right)\label{h0}\eeq
and
\beq {\cal{H}}^{(1)} = \left(\begin{array}{cc} -\frac{\gamma}{2} H_z\hat{\sigma}_z & \Delta_{0}(ik_y - k_x\Pz)/k_F \\ -\Delta_{0}(ik_y + k_x\Pz)/k_F &\frac{\gamma}{2} H_z\hat{\sigma}_z \end{array}\right).\label{h1}\eeq
We look for zero-energy eigensolutions of (\ref{h0}) of the form
\beq \hat{\phi}_i(z>0) = \frac{1}{\sqrt{A}}\hat{\alpha}_ie^{iKz},\label{phi}\eeq
where $\hat{\alpha}_i$ is a 4-component spinor, and $A$ is a normalization constant.  One thus finds the solutions
\beq \hat{\alpha}_1 = \frac{1}{\sqrt{2}}\left(\begin{array}{c} 1\\ 0 \\ 0 \\ -i\end{array}\right),\;\;\hat{\alpha}_2 = \frac{1}{\sqrt{2}}\left(\begin{array}{c} 0 \\ i\\ 1 \\ 0\end{array}\right),\label{eigenvectors0}\eeq
with $K = i\Delta_{0}/v_F$, corresponding to an exponential decaying solution, as appropriate for a bound surface state.  Note that (\ref{eigenvectors0}) are eigenvectors of the $z$-component of the spin operator, given by
\beq \hat{S}_z  = \left(\begin{array}{cc} \Pz & 0\\ 0 & -\Pz\end{array}\right),\label{Szdef}\eeq
with eigenvalues $\pm 1$.  

The eigenstates (\ref{phi}), (\ref{eigenvectors0}) of (\ref{h0}) will now be used to solve $({\cal{H}}^{(0)}+{\cal{H}}^{(1)})\hat{\Psi} = E\hat{\Psi}$ perturbatively.  Consider a solution of the form
\beq \hat{\Psi}_{\bk}(z) = \sum_{j=1}^2\hat{X}_j(\bk) \hat{\phi}_j(z).\label{eigenvectors} \eeq  
Using this, one finds that $\hat{X}(\bk)$ satisfies the two-dimensional Dirac equation
\beq \left[c_M\bk\cdot\vec{\mathbf{\sigma}} + \frac{\gamma}{2}H_z\Pz\right]\hat{X} = -E\hat{X},\label{Dirac}\eeq
where $c_M\equiv \Delta_{0}/k_F$.  Solving (\ref{Dirac}) for the normalized eigensolutions gives
\beq E^{\pm}_{\bk} = \pm \sqrt{(c_M\bk)^2 + (\gamma H_z/2)^2},\label{E}\eeq
with corresponding eigenvectors
\begin{align} X^{\pm}_1 &= \frac{c_M(k_x-ik_y)}{\sqrt{(c_M\bk)^2 + (E^{\pm}_{\bk} - (\gamma H_z/2))^2}},\nonumber\\
X^{\pm}_2 &= \frac{E^{\pm}_{\bk}-(\gamma H_z/2)}{\sqrt{(c_M\bk)^2 + (E^{\pm}_{\bk} - (\gamma H_z/2))^2}}.\label{eigenvectorsD}\end{align}

With (\ref{phi}),  (\ref{eigenvectors0}), and (\ref{eigenvectorsD}), (\ref{eigenvectors}) thus describes the in-gap Majorana states with dispersion (\ref{E}).  A gapless spectrum with characteristic zero-energy Majorana mode only arises in zero field, $H_z=0$.  We use these solutions to calculate the mean-field spin susceptibility per unit area 
\beq \chi^{(0)}_{zz}(q) = -\frac{1}{\beta}\int^{h}_0 dz \sum_{k}\mathrm{tr}[\bG(k+q;z)\hat{S}_z\bG(k;z)\hat{S}_z].\label{chizz0}\eeq
Because of the inhomogeneity along $z$, it will be convenient to always integrate over this axis.  All response functions in this paper will thus be those for a unit area.  The Green's functions entering (\ref{chizz0}) are the mean-field $4\times 4$ matrix Green's functions appropriate for the geometry described earlier, a superfluid confined to $0<z<h$.  It should thus include contributions not just from the low-energy surface modes, but also bulk continuum states.  As a simple ansatz, we use
\begin{align} \bG(k;z>0) &= \bG_{\mathrm{bulk}}(k)\Theta(h-z)\Theta(E_{b,\bk}-\Delta_0) \nonumber\\&+ \bG_{\mathrm{surface}}(k;z)\Theta(\Delta_0-E_{b,\bk}),\label{Gansatz}\end{align}
where
\beq \bG_{\mathrm{bulk}}^{-1}(k)\equiv \left(\begin{array}{cc}i\omega_n-\xi_{\bk}+\frac{\gamma}{2} H_z\hat{\sigma}_z& \hat{\Delta}(\bk)\\
\hat{\Delta}^{\dagger}(\bk)&i\omega_n + \xi_{\bk} -\frac{\gamma}{2} H_z\hat{\sigma}_z\end{array}\right)\label{Gbulk}\eeq
is the Green's function for bulk continuum excitations with $\xi_{\bk} = \bk^2/2m-\mu$ and where $E_{b,\bk}$ is the bulk BCS quasiparticle spectrum, given by the poles of $\bG_{\mathrm{bulk}}$.  $\Theta$ is the Heaviside function and the quasi-two-dimensional in-gap states are described by the surface Green's function
\beq \bG_{\mathrm{surface}}(k; z) = \frac{[\hat{\Psi}^{+}_{\bk}(z)]^{\dagger}\hat{\Psi}^{+}_{\bk}(z)}{i\omega_n - E^+_{\bk}} +  \frac{[\hat{\Psi}^{-}_{\bk}(z)]^{\dagger}\hat{\Psi}^{-}_{\bk}(z)}{i\omega_n - E^{-}_{\bk}},\label{Gsurface}\eeq
with $\pm$ referring to the $\pm$ branches (\ref{E}) and (\ref{eigenvectorsD}).   
In the above expressions, $k\equiv (\bk,i\omega_n)$ and $q\equiv (\bq,i\nu_m)$, where $\nu_m (\omega_n)$ are Bose (Fermi) Matsubara frequencies.  

The bulk susceptibility obtained from (\ref{Gbulk}) is purely real for $\omega<2\Delta_0$; the only contribution to the imaginary susceptibility at low frequencies ($\omega\lesssim \Delta_0$) thus comes from the surface contribution.  Using (\ref{Gsurface}) in (\ref{chizz0}), doubling to account for both surfaces and analytically continuing the external Bose frequency, $i\nu_m \to \omega+i0^+$, gives
\begin{align} &\mathrm{Im}\chi^{(0)}_{zz}(\b0,\omega) = \frac{\omega}{4c^2_{M}}\left(1-\frac{\omega^2_L}{\omega^2}\right) \tanh\left(\frac{\beta \omega}{4}\right)\Theta(\omega-\omega_L)\label{imchizz0}\end{align}
for the imaginary part of the zero-momentum limit of the spin susceptibility.  Here $\omega_L \equiv \gamma H_z$ is the Larmor frequency.

Apart from a factor of two accounting for the two surfaces in our geometry, the result shown by (\ref{imchizz0}) is equivalent to that derived by Silaev~\cite{Silaev11} [he deals with the \emph{magnetic} susceptibility $\chi_M = (\gamma^2/4)\chi_{zz}$, however].  It suggests a promising way to observe Majorana bound states at the surface of a thin film of \HeB using NMR which conventionally probes the long-wavelength $\bq\to \b0$ (i.e., uniform magnetic field) limit of the imaginary part of the susceptibility tensor~\cite{Chung09,Nagato09,Silaev11,Tsutsumi11,Mizushima12}.   At the mean-field level, only the Majorana branch provides an absorption channel for such a spin probe.  We expect collective spin modes to renormalize this response, however, analogous to the way collective density modes enter the optical Hall conductivity (\ref{sigmaHall}).  In the next section, we derive the low-energy theory of the in-plane longitudinal spin modes of a thin film of superfluid $^3$He-$B$.  The inclusion of these modes is crucial to calculating a spin susceptibility consistent with conservation laws (i.e., a locally gauge-invariant result), which we do in Sec.~\ref{GIspinresponse}.

\section{Longitudinal spin modes in a thin film of $^3$He-$B$}
\label{spinmodesec}
Following conventional notation (see the appendix), (\ref{hz}) can be written as $\hat{\Delta}(\bk) = \sum_{\alpha j}id_{\alpha j}k_{\alpha}(\Pj\Py)$, where the tensor $d_{\alpha i}$ is diagonal:  $d_{i\alpha} =  (\Delta_0/k_F)\delta_{\alpha i}$.   Spin excitations correspond to small rotations of the order parameter in spin space by $\vec{\theta} = (\theta_x,\theta_y,\theta_z)$~\cite{Brinkman74b}:
\beq d_{\alpha i} = \frac{\Delta_0}{k_F}\left(\begin{array}{ccc} 1 & \theta_z & -\theta_y\\ -\theta_z & 1 & \theta_x \\ \theta_y & -\theta_x & 1\end{array}\right).\label{OPfluct}\eeq

NMR probes the dynamics associated with the longitudinal spin mode, described by the phase angle $\theta_z$.  While in general the longitudinal spin dynamics are coupled to the transverse $(\theta_x,\theta_y)$ ones,  in the situation of interest to us, these modes decouple and we will be able to ignore the dynamics associated with $\theta_x$ and $\theta_y$.  First, in the limit where $\bq=\b0$, as long as the external magnetic field $\bH$ is sufficiently small ($H\lesssim 25$G) that the order parameter retains the thin-film form assumed above~\cite{Brinkman74}, the dipole force does not couple the transverse and longitudinal spin fluctuations~\cite{Osheroff74}.  Second, for $\bq\neq \b0$, gradient terms can also couple transverse and longitudinal spin modes [see, e.g., Ref.~\onlinecite{Brinkman74b}].  However, for the order parameter given by (\ref{OPfluct}) and restricting dynamics to lie in the plane, $\bq = (q_x,q_y,0)$, the coupling vanishes.  This restriction on $\bq$ is justified as long as the film thickness $h$ is not significantly larger than the coherence length $\xi_0\equiv v_F/\Delta_0$ since in this case the low-energy $\omega\ll \Delta_0$ collective modes of interest to us are frozen out along the $z$ axis: $v_Fq_z \sim 2\pi \Delta_0 (\xi_0/h)$.   In what follows, we will assume that transverse and longitudinal spin modes are decoupled, considering the situation where the magnetic field $H_z$ is small and the spin dynamics are in-plane.  Our treatment also assumes that the mean-field order parameter (\ref{hz}) is not \emph{qualitatively} renormalized by the coupling to spin dynamics; in Ref.~\onlinecite{Park14}, it is argued that the spin modes mediate an interaction between surface Majorana states that could destabilize this state.

Focussing only on the phase angle $\theta_z$ associated with the longitudinal spin mode in (\ref{OPfluct}), this small rotation amounts to the ``phase twists'' $\Delta_{\uparrow\uparrow} \to \Delta_{\uparrow\uparrow}\exp(-i\theta_z)$ and $\Delta_{\downarrow\downarrow} \to \Delta_{\downarrow\downarrow}\exp(i\theta_z)$ on the $\uparrow\uparrow$ and $\downarrow\downarrow$ components of the order parameter matrix (\ref{hz}).   Writing down the effective BCS action for this system, these phases can be ``gauged out'' by the transformation $\psi_{\uparrow}\to \psi_{\uparrow}\exp(i\theta_z/2)$,  $\psi_{\downarrow}\to \psi_{\downarrow}\exp(-i\theta_z/2)$~\cite{Aitchison95}.   We will also decompose the magnetic field $\bH(\br,t) = \mathbf{\hat{z}}[H_z + \delta H_z(\br,t)]$ into a background uniform, static field $H_z$ and the probe field $\delta H_z(\br,t)$ (as appropriate for longitudinal NMR, both are aligned along the $z$-axis), treating the latter as a perturbation.  Proceeding in the usual fashion to derive the Gaussian fluctuation action with respect to the phase~\cite{Aitchison95} and adding the dipole energy $H_D$, one obtains the following effective action (per unit area) for the spin phase degree of freedom in unpolarized~\cite{polarizationcomment} \HeBp:
\begin{align} S_{\mathrm{eff}}[\theta_z]&=\frac{1}{2}\sum_{q,\mu,\nu}\Big\{\left[\frac{n}{m}\delta_{\mu\nu}+\chi^{(0)}_{J^{s}_{\mu}J^{s}_{\nu}}(q)\right]A_{\mu}(q)A_{\nu}(-q)\nonumber\\&-\frac{\chi^{(0)}_{zz}(q)}{4}A_0(q)A_0(-q) \Big\} - H_D(\theta_z). \label{Seff} \end{align} 
Here, $q\equiv (\bq,\omega)$, $A_0(q)\equiv [\omega\theta(q)-\gamma  \delta H_z(q)]/2$, and $A_{\mu}(q)\equiv q_{\mu}\theta(q)/2$, where $\delta H_z(q)$ is the Fourier transform of the small magnetic probe field.  $\chi^{(0)}_{zz}(\bq,\omega)$ is the mean-field longitudinal susceptibility, given by (\ref{chizz0}), and $\chi^{(0)}_{J^{s}_{\mu}J^{s}_{\nu}}(\bq,\omega) = \beta^{-1}\int^h_0 dz\sum_k \mathrm{tr}[\bG(k;z)\hat{v}^{s}_{\mu}(\bk,\bk+\bq)\bG(k+q;z)v^{s}_{\nu}(\bk+\bq,\bk)]$ is the mean-field paramagnetic spin-current correlation function, where the spin velocity operator is 
\beq \hat{v}^s_{\mu}(\bk,\bk') \equiv \frac{(\bk+\bk')_{\mu}}{2m}\left(\begin{array}{cc}\Pz& 0 \\ 0 & \Pz\end{array}\right),\label{Vop}\eeq
and $n \equiv (\beta)^{-1}\int^{h}_0 dz \sum_k\mathrm{tr}[\bG(k;z)]$ is the number density per unit area.  

In (\ref{Seff}), $H_D$ is the dipolar energy density per unit area.  For small $\theta_z$ it can be expanded as~\cite{Brinkman74b}
\beq H_D \sim \mathrm{const.} + \frac{\chi_0 \Omega^2_{l}}{8}\theta^2_z,\label{HD}\eeq
where $\chi_0$ is the isotropic real susceptibility per unit area (equal to the usual susceptibility density multiplied by $h$) at zero wavelength and frequency.  $\Omega^2_l\propto \Delta^2_0(T)$ is the temperature- and pressure-dependent longitudinal resonance frequency in the $B$ phase measured by (zero wavelength) longitudinal NMR experiments~\cite{Leggett73,Osheroff74b,Osheroff74c}.  Unlike the transverse resonance in the vicinity of the Larmor frequency, the longitudinal resonance only arises in the superfluid phase, with $\Omega_l$ proportional to $\Delta_0(T)$.  For $T\sim 0.5T_c$, it is on the order of $\sim 2\pi\times 200$kHz~\cite{Osheroff74b}, well below the bulk gap ($\Omega_l \sim 10^{-2}\Delta_0$ for $\Delta_0\sim 1$mK).  

The spectrum $\omega_{\bq}$ of the spin mode probed in longitudinal NMR experiments is given by the poles  $\Gamma_{\theta_z}(\omega_{\bq},\bq) = 0$ of the fluctuation propagator obtained from the Gaussian action $S_{\mathrm{eff}}[\theta_z,\delta H_z=0] \equiv \tfrac{1}{2}\sum_{q}\theta_z(q)\Gamma^{-1}_{\theta_z}(q)\theta_z(-q)$.   As noted earlier, typical longitudinal NMR measurements on \HeB are sensitive to long-wavelength physics and hence probe the $\bq \to \b0$ limit of the collective modes where $\Gamma^{-1}_{\theta_z}(\omega,\b0) = \chi_0(\omega^2-\Omega^2_l)/4$ and a sharp resonance appears at $\omega= \Omega_l$~\cite{Osheroff74b,Osheroff74c}.

\section{Locally gauge-invariant spin susceptibility}
\label{GIspinresponse}
We now return to the problem of finding signatures of Majorana surface modes in the longitudinal susceptibility.  The gauge-invariant susceptibility can now be calculated as a functional derivative of the action (\ref{Seff}):
\begin{align} &\chi_{zz}(\bq,\omega) = \frac{4}{\gamma^2}\left.\frac{\partial^2}{\partial \delta H_z(q)\partial \delta H_z(-q)}\ln \int {\cal{D}}[\theta]e^{-S_{\mathrm{eff}}}\right|_{\delta H_z = 0}\nonumber\\ &=  \frac{\chi^{(0)}_{zz}(\bq,\omega)\left[\bq^2\gamma^2\bar{\chi}^{(0)}_{J^{s}J^{s}}(\bq,\omega)+\chi_0\Omega^2_l\right]}{\bq^2\gamma^2\bar{\chi}^{(0)}_{J^{s}J^{s}}(\bq,\omega)+\chi_0\Omega^2_l-\omega^2 \chi^{(0)}_{zz}(\bq,\omega)}.\label{chizz}\end{align}
Here we have defined $\bar{\chi}^{(0)}_{J^{s}J^{s}}\equiv [(\bar{n}/m)+\chi^{(0)}_{J^{s}_{\mu}J^{s}_{\nu}}]\delta_{\mu\nu}$ for this isotropic system; all correlation functions are calculated using (\ref{Gansatz}).   This result shows that when the dipole energy is zero, $\Omega_l\to 0$, the $\bq\to 0$ limit of the gauge-invariant longitudinal susceptibility vanishes, in contrast to the mean-field result (\ref{imchizz0}).  Since NMR probes the long-wavelength limit \beq \chi_{zz}(\b0,\omega) =  \chi^{(0)}_{zz}(\b0,\omega)\frac{\chi_0\Omega^2_l}{\chi_0\Omega^2_l-\omega^2 \chi^{(0)}_{zz}(\b0,\omega)}\label{chizzq0}\eeq
of the susceptibility, the NMR response (including that arising from Majorana modes) vanishes without the explicit inclusion of the dipole energy.   The other major feature of the gauge-invariant susceptibility is the presence of a pole at $\omega\sim \Omega_l$ describing the resonant absorption of longitudinal spin modes; i.e., the longitudinal resonance.  In the remainder of this section, we discuss the implications of this pole for the spin susceptibility measured in NMR experiments on thin films of \HeBp.  

For a film of thickness $h\gtrsim \xi_0 \simeq 10^3k^{-1}_F$, the real part of the mean-field susceptibility is dominated by the bulk contribution from (\ref{Gbulk}), whereas the imaginary part of the mean-field susceptibility at frequencies inside the bulk gap arises solely from the surface states.  Thus, at $T\ll E_F$ and frequencies well below $\Delta_0$, 
\begin{align}  \chi^{(0)}_{zz}(\b0,\omega) &\simeq 8h\sum_{\bk}\frac{\Delta^2_0(k^2_x+k^2_y)/k^2_F}{E_{b,\bk}(4E^2_{b,\bk}-\omega^2)} + i\mathrm{Im}\chi^{(0)}_{zz}(\b0,\omega)\nonumber\\
&\simeq \chi_0+  i\mathrm{Im}\chi^{(0)}_{zz}(\b0,\omega),\label{chitotal}\end{align}
where $\mathrm{Im}\chi^{(0)}_{zz}(\b0,\omega)$ is given by (\ref{imchizz0}) and $\chi_0 = 2hn/E_F$, with $n=k^3_F/3\pi^2$ the three-dimensional density.   Using $\xi_0\equiv v_F/\Delta_0$,  (\ref{imchizz0}) can be written as
\begin{align}\frac{\mathrm{Im}\chi^{(0)}_{zz}(\b0,\omega)}{\chi_0}  &= \frac{3\pi^2 }{16} \left(\frac{\omega}{\Delta_0}\right)\left(\frac{\xi_0}{h}\right)\tanh\left(\frac{\beta \omega}{4}\right)\nonumber\\&\times \left(1-\frac{\omega^2_L}{\omega^2}\right)\Theta(\omega-\omega_L).\label{imchizz0b}\end{align}
Not surprisingly, the relative magnitude of the absorption due to surface states as compared to the bulk susceptibility $\chi_0$ is proportional to the ratio $\xi_0/h$ of the coherence length and the film thickness.   Note that for $\omega\ll \Delta_0$, the tanh factor is also small since $T<T_c\sim \Delta_0/1.76$ (using the weak-coupling BCS result).  Without the absorption from Majorana surface states, the imaginary part of (\ref{chizzq0}) describes a Dirac delta function at the longitudinal resonance frequency $\omega=\Omega_l$ with weight $\pi \chi_0 \Omega^2_l/2$.  

\begin{center}
\begin{figure}
\includegraphics[width=0.8\linewidth]{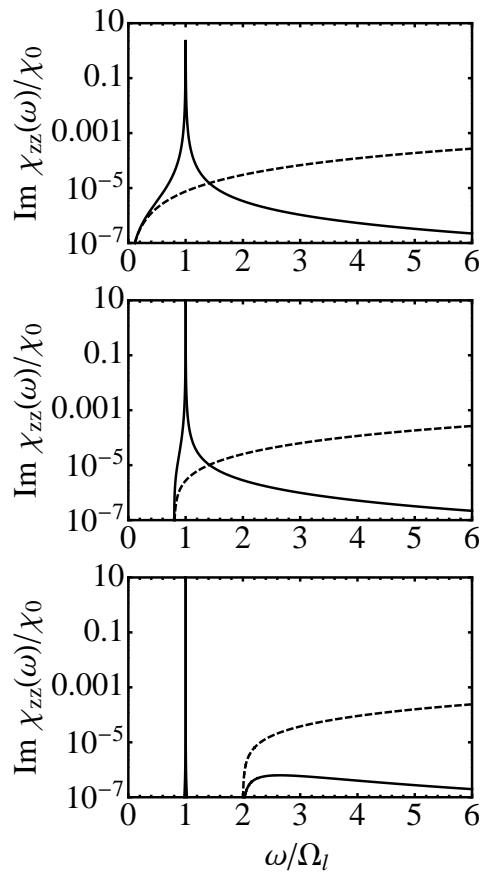}
\caption{Evolution of the gauge-invariant (solid line) and mean-field (dashed line) longitudinal spin susceptibilities with increasing magnetic field in a film of thickness $h=20\xi_0$ and $T=0.5T_c$ (using $T_c = \Delta_0/1.76$) with $\Omega_l = 2\pi \times 200\mathrm{kHz}$ and $\Delta_0 = 10^{-3}E_F$ (corresponding to $\Omega_l = 10^{-2}\Delta_0$). From top to bottom, $\omega_L/\Omega_l = 0, 0.8,$ and 2.  Increasing the magnetic field from zero leads to a gap in the spectrum of Majorana surface states and, for $\omega/\Omega_l<\omega_L/\Omega_l$, there is no absorption from these states.}\label{fig2}
\end{figure}
\end{center}

A key feature of Silaev's proposal~\cite{Silaev11} to detect Majorana surface states using longitudinal NMR lies with the magnetic field tunability of the mean-field response (\ref{imchizz0b}).  It vanishes at frequencies below the Larmor frequency $\omega_L$ as a result of the ``mass gap'' in the Majorana dispersion (\ref{E}) and hence, by tuning the external magnetic field, the presence of Majorana surface states can be discerned from the NMR spectrum.   This feature persists in the gauge-invariant result (\ref{chizzq0}), as can be seen in Fig.~\ref{fig2}, where we plot the imaginary part of (\ref{chizzq0}) using (\ref{chitotal}), as well as the mean-field result (\ref{imchizz0b}) for three different values of magnetic field.  In the top panel, there is no magnetic field and the spectral response is nonzero at all frequencies; in the gauge-invariant theory, it is strongly peaked at the longitudinal resonance at $\omega\simeq \Omega_l$.  With increasing magnetic field, the absorption due to Majorana surface states vanishes for all frequencies below $\omega_L$.  Once $\omega_L$ exceeds the longitudinal resonance frequency $\Omega_l$, the only spectral feature remaining below $\omega_L$ is the Dirac delta function at $\Omega_l$ (a small broadening has been added to the lowest panel in Fig.~\ref{fig2} to make the delta function visible), describing the resonant absorption of undamped spin modes.

Although both the mean-field and gauge-invariant responses evolve in a characteristic way with increasing magnetic field, in contrast to the mean-field expression, the spectral response of the gauge-invariant theory becomes \emph{strongly} suppressed well above the longitudinal resonance pole (note that Fig.~\ref{fig2} is a semi-log plot): $\mathrm{Im}\chi_{zz}(\omega\gg\Omega_l) \to  \mathrm{Im}\chi^{(0)}_{zz}(\omega)(\Omega_l/\omega)^4$.   The mean-field and gauge-invariant response functions coincide in the low-frequency limit, $\mathrm{Im}\chi_{zz}(\omega\ll\Omega_l) \to  \mathrm{Im}\chi^{(0)}_{zz}(\omega)$.  The response here is again very small, however,  due to the factors of $\omega/\Delta_0$ in (\ref{imchizz0b}) (recall that $\Omega_l \ll\Delta_0$ and hence, $\omega/\Delta_0 \ll \omega/\Omega_l$) and $\tanh(\beta \omega/4)$.

Despite the much smaller spectral response in the gauge-invariant theory away from the longitudinal resonance, it may still be possible to infer the existence of Majorana surface states from the magnetic-field tunability of the \emph{broadening} of the resonance arising from the contribution of $\mathrm{Im}\chi^{(0)}_{zz}(\b0,\omega)$ in the denominator of (\ref{chizzq0}).  Physically, this broadening arises from the decay (``Landau damping'') of a longitudinal spin excitation into a particle-hole pair on the Majorana branch~\cite{comment}:
\beq \omega_{\bq} = E^+_{\bk} - E^-_{\bk-\bq}.\label{Beliaev}\eeq
The broadening should be more pronounced in thinner films where (\ref{imchizz0b}) is larger.  We note that there is no analogue of this damping process for the chiral \emph{edge} modes of a two-dimensional chiral $p$-wave superfluid: The modes at a given edge belong to a single chiral branch, either $E_{k} = ck$ or $E_k=-ck$ (but not both), and (\ref{Beliaev}) cannot be satisfied.   It is only because 
the Majorana surface states are described by the two-dimensional Dirac cone (\ref{E}) that there is phase space available to satisfy this constraint.

\begin{center}
\begin{figure}
\includegraphics[width=0.8\linewidth]{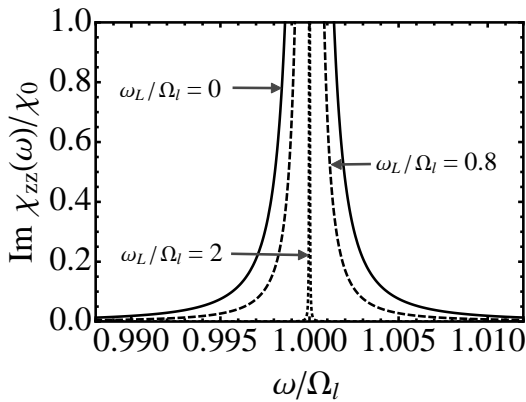}
\caption{The imaginary part of the gauge-invariant longitudinal susceptibility in the immediate vicinity of the longitudinal resonance pole for several values of the background magnetic field corresponding to different values of the Larmor frequency $\omega_L$.   The values of  $h$, $T$, $\Delta_0$, and magnetic fields are the same as those used in Fig.~\ref{fig2}.  A small broadening $\omega \to \omega + i10^{-9}$ has been added to make the resonance visible even when $\mathrm{Im}\chi^{(0)}_{zz} = 0$.}\label{fig3}
\end{figure}
\end{center}

The broadening is small on the scale of $\Omega_l\sim 10^3$kHz, and is not readily evident in Fig.~\ref{fig2}.  
In Fig.~\ref{fig3}, we plot the imaginary part of the susceptibility over a narrow range of frequencies about the longitudinal resonance pole at $T=0.5T_c$ for several magnetic fields, corresponding to the same values used in Fig.~\ref{fig2}. (At $T=0$, the linewidth is roughly ten times larger.)  For $\Omega_l\simeq \omega < \omega_L$, the broadening of the resonance pole vanishes.  Given the small size of the longitudinal resonance frequency, a modest magnetic field on the order of a Gauss would suffice to push the Larmor frequency above this threshold.   In this way, the spectrum (\ref{E}) of the Majorana branch can be probed by measuring the magnetic field dependence of the linewidth of the longitudinal NMR resonance.   Of course, the Majorana splitting (\ref{splitting}) must be kept smaller than the longitudinal resonance frequency to see this feature.  Using (\ref{splitting}) and $\Omega_l=10^{-2}\Delta_0$, this means that $h\gtrsim 15\xi_0$ (in all plots, we use $h = 20\xi_0$).

In closing this section, we note that (\ref{chizzq0}) satisfies the sum rule~\cite{Leggett72,Leggett73}
\beq \frac{2}{\pi}\int^{\infty}_0 d\omega \omega \mathrm{Im} \chi_{zz}(\b0,\omega) = 
 \chi_0\Omega^2_l
\label{sumrule}\eeq
 for the longitudinal spin susceptibility only when the bulk contribution to the mean-field susceptibility (\ref{chitotal}) is used to evaluate (\ref{chizzq0}).  This sum rule is strictly valid for a system with translational invariance, however, and it is likely that the breaking of this symmetry needed to have surface modes leads to a new term in the sum rule, perhaps related to the dipole-surface energy~\cite{Brinkman74}.  Without the dipole energy, the Hamiltonian commutes with $\hat{S}_z$ and, consistent with (\ref{chizzq0}), the $\bq\to \b0$ longitudinal spin susceptibility must vanish at all frequencies [in contrast to the mean-field result (\ref{imchizz0})].  This means that correct sum rule for the thin film geometry must also vanish as $\Omega_l\to 0$.

\section{Discussion}
\label{conclusions}

In this paper, we have derived a gauge-invariant theory of the longitudinal spin susceptibility of a thin film of \HeBp.  The inclusion of effects related to the dynamics of the superconducting order parameter leads to a strong renormalization of the spin response past its mean-field value.  The same physics arises in the Hall conductivity (\ref{sigmaHall}) of a chiral $p$-wave superconductor, which (\ref{chizzq0}) strongly resembles.  
Both results show that in order to discern the signature of the Majorana branch in two-particle response functions, one must probe the spectral region below and in the vicinity of the collective mode pole.  At frequencies above this pole, the response is greatly suppressed as compared to its mean-field value.  

Different than the \emph{bulk} Hall response, though, Majorana states at the surfaces of thin films of \HeB are also manifested as a broadening feature of the collective mode pole in the spin susceptibility.  This opens the door to detecting Majorana states by measuring the width of the longitudinal resonance in NMR.  (Progress towards probing films of \HeB down to $h\sim 10\xi_0$ using NMR is reported in Refs.~\onlinecite{Levitin13,Levitin14}.)  Notably, the broadening decreases with increasing background static magnetic field as a result of the growing gap in the Majorana spectrum, proportional to this magnetic field.  Because of the $\tanh (\beta \Omega_l/4)$ thermal factor suppressing the broadening of the longitudinal spin resonance and the fact that $\Omega_l\ll \Delta_0$, it will be challenging to observe this broadening.  Our calculations (shown in Fig.~\ref{fig3}) suggest that at $T\sim 0.5T_c$, the characteristic broadening will be ${\cal{O}}(10^{-3}\Omega_l)$, on the order of a few kHz.  Measurements of the \emph{transverse} resonance in thin films show linewidths $\lesssim 1$ kHz~\cite{Levitin13}.  In bulk systems, the longitudinal resonance linewidths are much broader than the transverse ones~\cite{Osheroff74b,Osheroff74c}.  This is likely due to textures in the order parameter, however, an effect which will likely be minimized in thin films.  

To conclude, although the low-energy dynamics of the superfluid order parameter obscures direct spectral signatures in the spin susceptibility of the in-gap Majorana states, the latter manifest themselves as a magnetic-field tunable damping of these dynamics, arising from the decay of a longitudinal spin mode into a  Majorana particle-hole pair.  As long as the background broadening of longitudinal resonance linewidths can be made comparable to those of the transverse resonance, longitudinal NMR experiments on thin films of \HeB should be able to find signatures of Majorana surface states in the absorption linewidth of the longitudinal resonance.

\acknowledgements
We thank Tony Leggett for alerting us to the fact that the mean-field susceptibility substantially violates the sum rule for the longitudinal spin susceptibility.   This work is supported by NSERC and CIFAR and by the Canada Research Chair and Canada Council Killam programs and the National Science Foundation under Grant No. NSF PHY11-25915 (CK). 

\appendix
\section{$^3$He-$B$ order parameter in a thin film}
\label{thinfilmOP}
For a system with translational invariance, the $^3$He order parameter can be written in terms of the tensor $d_{\alpha i}$ as~\cite{Leggett75,VWbook}
\beq\hat{\Delta}(\bk) = \sum_{\alpha j}id_{\alpha j}k_{\alpha}(\Pj\Py).\label{OP0}\eeq
The order parameter is a $2\times 2$ matrix in spin space, where e.g., the component that represents pairing between atoms with nuclear spins $\beta$ and $\gamma$ is given by $(\hat{\Delta})_{\beta\gamma}$.  
$d_{\alpha i}$ couples spin and orbital degrees of freedom, with $k_{\alpha}$ the $\alpha$ component of the momentum $\bk$.  One possible choice for the  \HeB order parameter is the diagonal tensor $d_{\alpha i} = (\Delta_0/k_F)\delta_{\alpha i}$.  Ignoring dipolar and surface effects, any relative rotation of spin and orbital axes yields a state degenerate with this one.  As a result, the  \HeB  order parameter is usually expressed as a rotation matrix $R_{\alpha i}$:
\beq d_{\alpha i} = (\Delta_0/k_F)R_{\alpha i} \delta_{\alpha i}.\label{OPdiag}\eeq

The rotation matrix is fully characterized by specifying an axis of rotation $\mathbf{\hat{n}}$ and an angle of rotation $\theta$ about this axis.  
While these are arbitrary for an infinite, uniform superfluid when the dipolar energy is ignored, in conjunction with surface effects, the dipolar energy fixes both of these quantities in a confined superfluid.  Even in a completely uniform system without boundaries, the dipolar energy fixes the angle of rotation, given by the so-called Leggett angle $\theta_L = \cos^{-1}(1/4)$~\cite{Leggett75}.  Including surface effects, the rotation angle is unchanged from $\theta_L$~\cite{Brinkman74}, while the axis of rotation $\mathbf{\hat{n}}$ is determined by the interplay between surface and dipolar effects and the external magnetic field~\cite{Leggett73,Engelsberg74,Brinkman74}, if any.  For sufficiently small magnetic field, $H\lesssim 25$G~\cite{Brinkman74} (see also, Ref.~\onlinecite{VWbook}, above 6.132) however, surface effects dominate and near the surface, the dipolar energy orients $\mathbf{\hat{n}}$ to be normal to the surface.  As long as the film is thinner than the dipole coherence length ($\sim 10\mu$m $\sim 100\xi_0$, where $\xi_0$ is the BCS coherence length), the axis of rotation is fixed to be normal to the surface throughout the film. 

Putting the above results together, the  \HeB order parameter in a thin film ($d\lesssim 100\xi_0$) and small magnetic field ($H\lesssim 25$G) is, in the basis spanned by $x,y,z$ Cartesian coordinates,
\beq d_{\alpha i} = \frac{\Delta_0}{k_F}\left(\begin{array}{ccc} \cos \theta_L & -\sin \theta_L & 0\\ \sin \theta_L & \cos\theta_L & 0 \\ 0 & 0 & 1\end{array}\right).\label{OP}\eeq
It is convenient to calculate quantities in a basis where the order parameter matrix is diagonal.  This is accomplished by fixing the spin axes and rotating the orbital ones in the $x-y$ plane by the Leggett angle.  For the calculations in this paper, we thus take the order parameter to be diagonal, given by (\ref{OPdiag}), but where it is understood that the orbital axes are rotated in the $x-y$ plane.  Combining (\ref{OP0}) and (\ref{OP}) gives 
\beq \hat{\Delta}(\bk)= \frac{\Delta_0}{k_F}\left(\begin{array}{cc} -k_x+ik_y & k_z\\ k_z & k_x+ik_y\end{array}\right).\eeq
In the semiclassical approximation for the situation where there is a surface at $z=0$, treating the order parameter amplitudes as constant above for $z>0$, this reduces to  (\ref{hz}) in the main text.


\begin{thebibliography}{99}
\bibitem{polarizationcomment} For a polarized superfluid, there would be an additional term $\propto \sum_q  \langle \hat{S}_z\rangle A_0(q)$ in the action that is linear in $A_0(q)$.
\bibitem{comment} At finite $T$, there is also a contribution to broadening from thermally excited continuum states.  At the low temperatures we consider, this contribution will be very small.  Moreover, it will not exhibit the same magnetic-field dependence as the damping from the Majorana branch.
\end{thebibliography}
\end{document}